\documentclass[prx,twocolumn,amsart]{revtex4-2}
\usepackage{graphics}
\usepackage{amssymb}
\usepackage{amsmath}
\usepackage{slashed}
\usepackage{epsfig}
\usepackage{color}
\usepackage{verbatim}
\usepackage{float}
\usepackage{mathrsfs}
\usepackage{esint}
\usepackage{multirow}
\usepackage{upgreek}
\usepackage{xcolor}
\usepackage[perpage]{footmisc}

\usepackage[LGR,T1]{fontenc}

\usepackage{tabularx}

\let\Right\right
\let\Left\left
\makeatletter
\def\right#1{\Right#1\@ifnextchar){\!\right}{}}
\def\left#1{\Left#1\@ifnextchar({\!\left}{}}
\makeatother

\usepackage[
            pdfstartview=FitH,
            bookmarksnumbered=true,
            bookmarksopen=true,
            colorlinks,
            linkcolor=blue,
            anchorcolor=green,
            citecolor=blue
            ]{hyperref}
\begin{document}

  \renewcommand\arraystretch{2}
 \newcommand{\bq}{\begin{equation}}
 \newcommand{\eq}{\end{equation}}
 \newcommand{\bqn}{\begin{eqnarray}}
 \newcommand{\eqn}{\end{eqnarray}}
 \newcommand{\nb}{\nonumber}
 \newcommand{\lb}{\label}
 
\newcommand{\La}{\Lambda}
\newcommand{\va}{\scriptscriptstyle}
\newcommand{\be}{\nopagebreak[3]\begin{equation}}
\newcommand{\ee}{\end{equation}}

\newcommand{\ba}{\nopagebreak[3]\begin{eqnarray}}
\newcommand{\ea}{\end{eqnarray}}

\newcommand{\la}{\label}
\newcommand{\n}{\nonumber}
\newcommand{\su}{\mathfrak{su}}
\newcommand{\SU}{\mathrm{SU}}
\newcommand{\U}{\mathrm{U}}
\newcommand{\red}{ }

\newcommand{\R}{\mathbb{R}}

 \newcommand{\cb}{\color{blue}}
    \newcommand{\cc}{\color{cyan}}
        \newcommand{\cm}{\color{magenta}}
\newcommand{\rc}{\rho^{\scriptscriptstyle{\mathrm{I}}}_c}
\newcommand{\rd}{\rho^{\scriptscriptstyle{\mathrm{II}}}_c} 
\NewDocumentCommand{\evalat}{sO{\big}mm}{%
  \IfBooleanTF{#1}
   {\mleft. #3 \mright|_{#4}}
   {#3#2|_{#4}}%
}
\newcommand{\PRL}{Phys. Rev. Lett.}
\newcommand{\PL}{Phys. Lett.}
\newcommand{\PR}{Phys. Rev.}
\newcommand{\CQG}{Class. Quantum Grav.}

\title{Inner Radius and Energy Conditions of Dark Matter Halos Surrounding Schwarzschild Black Holes}
 
\author{Zibo Shen${}^{a,b}$}
\email{zshen284@wisc.edu}

\author{Anzhong Wang${}^{a}$}
\email{anzhong$\_$wang@baylor.edu}

\author{Shaoyu Yin${}^{c}$}
\email{syyin@zjut.edu.cn; the corresponding author}
 
\affiliation{${}^{a}$ GCAP-CASPER, Department of Physics and Astronomy, Baylor University, Waco, Texas 76798-7316, USA\\
${}^{b}$ Department of physics, University of Wisconsin-Madison, Madison, Wisconsin, USA\\
${}^{c}$ Institute for Theoretical Physics and Cosmology, Zhejiang University of Technology, 
Hangzhou 310023, China.}

\begin{abstract}

We study a class of analytic models for a dark matter halo surrounding a Schwarzschild black hole sitting at the center of a galaxy, with a variable inner radius $r_{\text{in}}$ at which the density profile of the dark matter halo vanishes. We examine in detail how the three energy conditions are satisfied in such models. In particular, independent of concrete profile, we find that the three energy conditions are satisfied when $r_{\text{in}}\ge5M/2$, where $M$ denotes the mass of the black hole. This indicates it is crucial to include inner radius when discussing dark matter distributions. All our solutions expressed explicitly in closed form are particularly valuable for the studies of the gravitational waveforms of extreme/intermediate mass ratio inspirals and the nature of dark matter in galaxies. 

\end{abstract}

\maketitle

\section{Introduction}\lb{sec:Intro}

Dark matter (DM) remains one of the key unsolved fundamental problems in physics \cite{Bertone:2018}. During the past decades since its gravitational effect was first observed~\cite{Zwicky:1933,Babcock:1939}, evidence of its extensive existence in the universe has accumulated from various aspects, such as the rotation curves \cite{Roberts:1973,Persic:1996}, the anisotropy in cosmic microwave background \cite{Hu:2002,Bennett:2013}, baryonic acoustic oscillation \cite{Eisenstein:2005}, gravitational lensing \cite{Clowe:2006}, etc. DM is found especially abundant in galaxies, forming halos surrounding the supermassive black holes (BHs) typically resting at each galactic center~\cite{Wang:2020}. It is believed to play significant roles in the evolution of the universe, the formation of large scale structures, and the dynamics of galaxies and their clusters~\cite{Aghanim:2020,Young:2017}. However, except for its gravitational effect, our knowledge about DM remains embarrassingly deficient. Nothing is certain about its microscopic constitution. Despite ingenious and elaborate searches in colliders and detectors based on scattering mechanism, till now experimentalists rule out disappointingly larger parameter spaces for various theoretical attempts \cite{DMexperiment}. What is certain about the DM property is that it takes part in gravitational interaction, but very likely isolates from other fundamental interactions, therefore the gravitational effect remains one of the most reliable approaches for the study of DM.

The epochal success in gravitational wave (GW) observation opens a new channel to examine the gravitational effects of DM which are hardly accessible by conventional means \cite{Cai:2017,Bailes:2021}. The extreme/intermediate mass ratio inspirals (EMRIs/IMRIs) \cite{Amaro-Seoane:2007}, for example, become highly attractive objects recently. An EMRI/IMRI system consists of a central supermassive BH with a much smaller mass orbiting around it, such that the dynamics can be studied analytically by perturbations. Typical EMRIs/IMRIs emit GWs with frequencies suitably in the sensitive range of the space-based GW detectors such as DECIGO~\cite{DECIGO}, LISA~\cite{LISA,Babak:2017}, Taiji~\cite{Taiji,Ruan:2018}, and TianQin~\cite{TianQin}, promoting EMRIs/IMRIs among the most promising potential sources for future GW observations. If the central BH is surrounded by a DM halo, which is very likely as observed in most galaxies, the GW of the corresponding EMRIs/IMRIs will be affected by the DM, in other words, the GW will carry information encoding the detailed  properties of the DM halo. Thus the GW observation will help to reveal the nature of DM. For this purpose, it is essential to be able to calculate the waveforms with which the effects of DM can be distinguished~\cite{Kavanagh:2020,Zhang:2024,Dai:2023}, thus analytic models are in great demand. 

To this end, five analytical models representing supermassive BHs surrounded by DM halos have been constructed recently \cite{Shen:2024}, including an important factor to restrict the density profile of the DM halo to vanish for $r\leq4M$. Since this requirement is based on full relativistic study of the standard Einstein theory \cite{Sadeghian:2013}, while the concrete DM properties, especially adjacent to the BH horizon, remain elusive, for the sake of a more flexible theory it is useful to generalize to cases with vanishing distributions inside arbitrary inner radius $r_{\text{in}}$. This also provides another advantage to examine the energy conditions. As has been pointed out in Ref.~\cite{Shen:2024}, all the five analytic models with inner boundary set at $4M$ satisfy the three energy conditions. It is tantalizing to ask what if the inner boundary location is set free, and what would be the case in other models. Actually, it was recently observed that some DM profiles without inner boundary violate the dominant energy condition very close to the BH horizon \cite{Datta:2024}. Since many recent works were based on DM profiles without inner boundary, such examination on energy conditions is timely and indispensable.

The generalization to arbitrary $r_{\text{in}}\geq 2M$ is mathematically straightforward based on previous results with $r_{\text{in}}=4M$, therefore, we shall present in this Letter only the extend Model I of Ref.~\cite{Shen:2024} but with more details as complement to the previous concise derivation. Then the energy conditions will be examined carefully under such generalization, but not limited to this Model I. In particular, we find a model-independent conclusion that the three energy conditions are satisfied when $r_{\text{in}}\ge5M/2$.

\section{Generalization to arbitrary inner boundary}\lb{sec:Model}

We start from the simplest case with a Schwarzschild BH surrounded by a halo in the form of Einstein cluster~\cite{Einstein:1939}. In such configurations with spherical static geometry, the massive DM particles move around the central BH in circular orbits where the role of centripetal force is played by the only interaction DM particles can feel: gravitation, which is always towards the center, even considering the massive halo, due to the spherical symmetry. In this way the halo demonstrates no radial pressure, while the angular components of the pressure, due to randomly distributed tangental velocity on every circular orbit, are isotropic functions of radius only. Admittedly, this is a rather simplified assumption, but it is self-consistent and permits stable configurations \cite{Geralico:2012}. More importantly, it yields analytic solutions if the density profiles of the DM halo are properly chosen \cite{Shen:2024}.

In above setup, with the standard 4d spherical coordinates $\{t,r,\theta,\phi\}$, the halo has energy-momentum tensor 
\bq
\lb{EMT}
T^\mu_\nu={\rm diag}(-\rho(r),0,P(r),P(r)),
\eq
where $\rho(r)$ is the energy density profile and $P(r)$ the tangential pressure, while the space-time metric can be written as 
\bq
\lb{metric}
ds^2=-f(r)dt^2+\frac{dr^2}{1-\frac{2m(r)}{r}} +r^2(d\theta^2+\sin^2\theta d\phi^2),
\eq
where $f(r)$ and $m(r)$ are functions to be determined by the Einstein equations \cite{Cardoso:2022}. In the present coordinates, one can finally get three coupled equations: 
\bqn
m'(r)=4\pi Gr^2\rho(r),\lb{eqm}\\
\frac{f'(r)}{f(r)}=\frac{2m(r)}{r[r - 2m(r)]},\lb{eqf}\\
P(r)=\frac{m(r)\rho(r)}{2[r-2m(r)]}.\lb{eqP}
\eqn
Eqs.~(\ref{eqm}) and (\ref{eqf}) can be solved by integration given suitable density profiles $\rho(r)$. Based on our previous work~\cite{Shen:2024}, we consider a more general form of $\rho(r)$ by relaxing the requirement that the halo vanishes at $r\le 4M$ where $M$ is the BH mass. In this work the inner boundary of DM distribution is set at $\upsilon M$
instead of $4M$, with an arbitrary real number $\upsilon$. Since the DM halo cannot exist inside the BH, we assume $\upsilon \geq 2$. Now we have 
\bq
\lb{eq2}
\rho(r)=\Theta(r-\upsilon M)\frac{\rho_0(1-\upsilon M/r)^n}{(r/a)^\gamma[1+(r/a)^\alpha]^{(\beta-\gamma)/\alpha}},
\eq
where $\Theta$ is the Heaviside step function, $\rho_0$ is a characteristic density, and $a$ is the characteristic scale of the halo, while $(n,\alpha,\beta,\gamma)$ are dimensionless parameters determining the shape of the density profile.

Actually, it has been shown that the inner boundary of the DM halos can be as small as $4M$ when full relativistic considerations are taken into account \cite{Sadeghian:2013,Speeney:2022}. However, in these analyses DM particles were allowed to have non-circular motions. In the current Letter, on the other hand, we assume that such radial motions are negligible. Though in many cases our simplification may no longer be valid, the analytic expressions can still be useful as an approximation for comparison with numerical results of other general models, if exact analytic solutions are unavailable for the latter. Since the gravitational effect due the DM halo is stronger when closer to the BH where its density peaks, and theoretically suggested DM annihilating processes are also expected to be more intensive there, the observational effects should be very sensitive to the inner distribution of DM. Hence, the current generalization of the previous models presented in Ref.~\cite{Shen:2024} will allow us to explore these effects, and whereby help us to distinguish different theoretical models through observations. 

The special profile we study here as an example is with $(n,\alpha,\beta,\gamma)=(1,1,4,1)$ (Model I in Ref.~\cite{Shen:2024}), i.e. 
\bq
\lb{rho}
\rho(r)=\rho_0\frac{\Theta(r-\upsilon M)(1-\upsilon M/r)}{(r/a)(1+r/a)^3}, 
\eq
which is of special interest as it coincides with the famous Hernquist model except for the additional factor $(1-\upsilon M/r)$ imposing the inner boundary. Integration of Eq.~(\ref{eqm}) with this $\rho(r)$ yields
\bq\lb{resm}
m(r)=M+\Theta(r-\upsilon M)\frac{(r-\upsilon M)^2}{(a+r)^2}M_h,
\eq
with the total mass of the halo $M_h=\frac{2a^4\pi\rho_0}{a+\upsilon M}$.

To solve $f(r)$, we make the replacement $x = r+a$ and use Eq.~(\ref{resm}) to rewrite Eq.~(\ref{eqf}) as
\bq
\lb{eq8}
\frac{f'(r)}{f(r)}= -\frac{1}{x-a}+\frac{x^2}{{\cal D}(x)},
\eq
where, as a cubic polynomial in terms of $x$,
\bqn
{\cal D}(x) &\equiv& x^3+Ax^2+Bx+C,
\eqn
with $A \equiv -(a+2M+2M_h)$, $B\equiv 8\pi a^4\rho_0=4M_h(a+\upsilon M)$, and $C\equiv -2M_h(a+\upsilon M)^2$. Its discriminant $\Delta \equiv \frac{p^3}{27} +\frac{q^2}{4}$ is positive, where $p\equiv B-A^2/3$ and $q\equiv 2A^3/27-AB/3+C$, because for the physically relevant cases in which $a\gg M_h\gg M$, we have $\Delta \simeq 2Ma^5_h/27>0$ to the leading order. Thus ${\cal D}(x) = 0$ has one real root (denoted by $x_1$) and a pair of complex conjugate roots (denoted by $x_{2, 3}$):
\bqn
\lb{roots}
x_1 &=& X_+ + X_--\frac{A}{3},\nb\\
x_2 &=& {\rm e}^\frac{2i\pi}{3}X_+-{\rm e}^{-\frac{2i\pi}{3}}X_--\frac{A}{3}=x_3^*,
\eqn
where 
\bqn
X_\pm \equiv \sqrt[3]{\frac{-q\pm\sqrt{\Delta}}{2}}.
\eqn 

Then, we can further split the term with ${\cal D}(x)$ to get 
\bq\la{feq}
\frac{f'(r)}{f(r)}=-\frac{1}{x-a}+\frac{\alpha_1}{x-x_1}+\frac{\alpha_2x+\alpha_3}{x^2+2kx+|x_2|^2},
\eq
where 
\bqn
\alpha_1&\equiv& \frac{x_1^2}{x_1^2+|x_2|^2+2k x_1}, \quad \alpha_2\equiv1-\alpha_1, \nb\\
\alpha_3&\equiv& \frac{|x_2|^2}{x_1}\alpha_1,\quad
k \equiv \frac{X_++X_-}{2}-\frac{A}{3}.
\eqn
Expression in the form of Eq.~(\ref{feq}) can lead to explicit integration \cite{Gradshtein:2007}. Setting the boundary condition of asymptotic flatness, $f(r\rightarrow\infty)=1$, we get
\begin{align}\lb{fr}
f(r)=&\frac{1}{r} (a+r-x_1)^{\alpha_1}\left[\left(a+r+k\right)^2+|x_2|^2\right]^{(1-\alpha_1)/2} \nb\\
&\times\exp\left(\frac{k\alpha_2-\alpha_3}{\sqrt{|x_2|^2-k^2}}{\rm arccot}\frac{a+r+k}{\sqrt{|x_2|^2-k^2}}\right),
\end{align}
for $r>\upsilon M$, which must connect continuously to a metric of the Schwarzschild solution with $f = f_0(1-\frac{2M}{r})$ in the vacuum region  $r\leq \upsilon M$, thus we find $f_0=f(\upsilon M)/(1-2/\upsilon)$. Hence, the space-time geometry is completely determined over the whole range 
$r \in (0, \infty)$.

As can be seen from comparison with Ref.~\cite{Shen:2024}, the results with $r_{\text{in}}=\upsilon M$ are just the replacement with $4M$ by $\upsilon M$, since this generalization merely modifies the lower bound of mathematical integration. This simple rule can be directly applied to other models given $\upsilon$ in a reasonable range, i.e., $2M\leq\upsilon M\ll M_h$, so we need not repeat the process for other models.

\section{Energy conditions}

It is generally accepted that any realistic solution to the Einstein equations should obey the three energy conditions, namely the weak, the strong, and the dominant energy conditions \cite{Hawking:2023}. These conditions are also closely related to singularity theorems and BH physics, as well as the causal structure of general spacetimes. The energy conditions can play vital roles in ruling out solutions to Einstein's field equations. Recently, for example, they have been used in examining the dark halo models with and without radial pressure \cite{Datta:2024}. Below we examine the energy conditions in our setup with variable $r_{\text{in}}$.

We start with the model solved in previous section. With the static spherically symmetric diagonal metric, the energy conditions for the Einstein cluster model reduce to: (1) $\rho(r)\ge0$ and $\rho(r)+P(r)\ge0$ for the weak energy condition (WEC); (2) $\rho(r)\ge0$ and $\rho(r)+2P(r)\ge0$ for the strong energy condition (SEC); and (3) $\rho(r)\ge|P(r)|$ for the dominant energy condition (DEC) \cite{Wald:1984}. 
In our models, we have obviously $\rho(r)\ge0$ and $P(r)\ge 0$ for $ r \geq \upsilon M$, so the WEC and SEC are always satisfied. Now we need only to consider the DEC, which reduces to $\rho(r)\ge P(r)$, or simply  
\bq
\lb{CD}
\frac{m(r)}{2{\cal{G}}(r)} \leq 1,
\eq
by using Eq.~(\ref{eqP}), where  ${\cal{G}}(r) \equiv r - 2m(r)$. To study the above condition,  let us first show that  ${\cal{G}}(r)$ is always a monotonically increasing function. Using the solution for $m(r)$ in Eq.~(\ref{resm}), we find 
\bqn\lb{Gp}
{\cal{G}}'(r) &=& \frac{1}{(a+r)^3}\big\{a^3+r\left[3a^2-4M_h(a+\upsilon M)\right]+3ar^2\nb\\
&&~~~~~~~~~~~~~ +4\upsilon M M_h(a+\upsilon M) +r^3\big\},
\eqn
which is always positive for $a\gg M_h\gg M$. Therefore, ${\cal{G}}(r)$ is a monotonically increasing function of $r$. At the inner radius $r = r_{\text{in}} (\equiv \upsilon M)$, this quantity takes its minimum as ${\cal{G}}(r_{\text{in}})=(\upsilon-2)M \geq0$, thus ${\cal{G}}(r) = r-2m(r)$ is guaranteed to be always non-negative and monotonically increasing. Besides, from Eq.~(\ref{eqf}) we also get that $f'(r)/f(r) > 0$, which tells us that $f(r)$ is also an increasing function of $r$, as $f(r)$ is always positive for $r > r_{\text{in}}$.

Note that when $\upsilon = 2$, that is, $r_{\text{in}} = 2M$, we find $m(r_{\text{in}}) = M$ and ${\cal{G}}(r_{\text{in}}) = 0$, indicating that the condition in Eq.~(\ref{CD}) cannot be satisfied near the BH horizon and completely breaks down on the horizon. Therefore, in the following  we only need to consider the case $\upsilon > 2$. Then, we find ${\cal{G}}(r)$ is always positive for $r \in [r_{\text{in}}, \infty)$. As a result, Eq.(\ref{CD}) is equivalent to 
\bq
\lb{CDb}
{\cal{F}}(r) \equiv 2r - 5 m(r)  \geq 0.
\eq
To study this condition, we first note that ${\cal{F}}(r)$ is a monotonically increasing function of $r$, since
\bqn\lb{Fp}
{\cal{F}}'(r) &=& 
  \frac{2}{(a+r)^3}\big\{a^3+r\left[3a^2-5M_h(a+\upsilon M)\right]+3a r^2\nb\\
  &&~~~~~~~~~~~~~ +5\upsilon MM_h(a+\upsilon M)+r^3\big\}>0,
\eqn
for $a\gg M_h\gg M$. So, at the boundary $r = r_{\text{in}}$, the function ${\cal{F}}(r)$ takes its minimal value 
\bq
\lb{MF}
{\cal{F}}_{\text{min}} \equiv   {\cal{F}}(r_{\text{in}}) = (2\upsilon - 5)M.
\eq
[Recall that $m(r_{\text{in}}) = M$]. Thus, the condition in Eq.~(\ref{CDb}) is satisfied for $r \in [r_{\text{in}}, \infty)$, provided that
\bq
\lb{CDc}
 \upsilon \geq \frac{5}{2}. 
\eq

Above derivation involves the analytic formulae obtained before, but the validity of its main conclusion can actually be generalized to many other models even without analytic solutions. As we can see from above steps, the criterion depends on the sign and slope of two terms ${\cal{G}}(r)$ and ${\cal{F}}(r)$, which are universal for Einstein cluster but independent of density profiles. Actually ${\cal{G}}(r)\equiv r-2m(r)>0$ can be generally inferred from the metric since the $dr^2$ term must remain positive outside the horizon to keep radial vectors spacelike, and the existence of DM halo cannot change this property. We can further claim that $r$ should be significantly larger than $2m(r)$ given that $r=2M$ describe the horizon of Schwarzschild BH, as wherever $r=2m(r)$ it is equivalent to form a Schwarzschild horizon there. Since the DM halo is much dilute compared with the central BH, we expect $2m(r)$ to be even smaller than $r$ as the radius increases, as shown explicitly by Eq.~(\ref{Gp}) in the concrete model. Similarly we can safely say that the slope of ${\cal{F}}(r)\equiv2r-5m(r)=2{\cal{G}}-m(r)$ should also be positive since the additional contribution from $m(r)$ is by all means not enough to balance the increment in $\cal{G}$ for a reasonable DM halo. In other words, the expressions concerned are much larger than zero when the inequalities are valid, so it does not matter whether the ratio between $r$ and $m(r)$ is $2$ or $2.5$. Actually, the angular momentum calculation in Ref.~\cite{Maeda:2024} shows explicitly that $r>3m(r)$, thus the requirement in Eq.~(\ref{CDb}) is always satisfied. In the model above, we can see by comparing Eq.~(\ref{Fp}) with Eq.~(\ref{Gp}) that the only negative term in ${\cal{F}}'(r)$ is only slightly larger than that in ${\cal{G}}'(r)$ due to the contribution from the extra weight of $m(r)$, but the whole combination remains much larger than $0$. Consequently, the most likely place for the inequality to fail is at smallest possible $r$, where the DM distribution is closest to the BH, which is actually $r_{\text{in}}$. Therefore, the criterion in Eqs.~(\ref{MF}) and (\ref{CDc}) applies to any reasonable DM halo profile. Interestingly, this conclusion is partially supported by the results in Ref.~\cite{Datta:2024}, where the $P/\rho$ curves for both Einasto and Hernquist profiles seem to cross over $1$ at roughly the same radius.

Finally, we want to point out that the lower bound for $r_{\text{in}}$ is rather small, given that for pure Schwarzschild BH, the time-like innermost stable circular orbit (ISCO) has radius $6M$, and even unstable circular timelike orbit has radius larger than $3M$. To examine the ISCO in the presence of DM halo of the Einstein cluster form, one can study the geodesic equation of time-like circular orbit with the metric of Eq.~(\ref{metric}). It is well known that the 4-velocity $u^\mu=(\partial/\partial\tau)^\mu$ for such an orbit is related to three conserved quantities, namely the energy $E$, angular momentum $L$ and its z-component $L_z$ (all per unit mass), as $u_0=-E$, $u_\theta^2=L^2-L_z^2/\sin^2\theta$ and $u_\phi=L_z$. With these one gets the radial equation $f(r)(dr/d\tau)^2/[1-2m(r)/r]=E^2-V^2_\mathrm{eff}$ with the effective potential $V^2_\mathrm{eff}=f(r)(1+L^2/r^2)$, thus the ISCO is determined by the condition of $dV^2_\mathrm{eff}/dr=d^2V^2_\mathrm{eff}/dr^2=0$, resulting in the simple equation
\bqn
\lb{eqISCO}
r^2m'(r) + m(r)\left[r -6m(r)\right] =0,
\eqn
which depends on the form of profile. For our model with $m(r)$ given in Eq.~(\ref{resm}), its derivative has denominator dominated by very large $a$ for typical halos, thus to solve the ISCO radius of the order of $M$, the first term in Eq.~(\ref{eqISCO}) is negligible and $m(r)\approx M$, yielding $r=6M$, which reduces to the pure Schwarzschild case. Keeping higher order terms, we find the halo effect appears at the order of $a^{-2}$, as $r\approx6M[1-MM_h(36-\upsilon^2)/a^2]$, implying that the existence of halo causes the ISCO slightly closer to the BH as long as $\upsilon<6$. A more general examination considering different shapes of $m(r)$ from various models found that the ISCOs could lie between $3M$ to $6M$ \cite{Maeda:2024}. The inner boundary $2.5M$ is smaller than all of above radii, however, this bound is still informative as the DM property is largely unknown yet. Besides, it can also be taken as a caveat when a DM halo profile without inner boundary is used in research. This concern could be significant because close to the BH horizon usually lies the peak density of DM where its local gravitational (or even collisional, if any) effects should be strongest.

\section{Conclusions and Discussions}

We have extended Model I found recently in Ref.~\cite{Shen:2024} to the case where the inner radius $r_{\text{in}}$ of the DM halo can take any real value, as long as it is no less than the Schwarzschild radius $r_s = 2M$ of the supermassive BH sitting at the center of a galaxy. When $r_{\text{in}} = r_s$, the solutions reduce to the one found recently in Ref.~\cite{Cardoso22}. However, in order for the DM halo to satisfy the three energy conditions, we find that the inner radius $r_{\text{in}}$ of the DM halo must satisfies the model-independent condition
\bq
r_{\text{in}} \geq \frac{5M}{2},
\eq
in the Einstein cluster configuration. This is consistent with the results found in Ref.~\cite{Cardoso22}, in which the authors showed that their model does not satisfy the dominate energy condition near the horizon $r \gtrsim 2M$.

It must be noted that physically our analytic solutions do represent spacetimes produced by different physical DM halos. These differences in $r_{\text{in}}$ should be also reflected to the gravitational waveforms emitted by compact objects orbiting around the supermassive BH. Clearly, depending on the ratio of the mass of the supermassive BH and the mass of the compact object, these GWs can be considered as emitted by either EMRI  or IMRI systems, which are the primary sources of GWs to be detected by the next generation of both space-based and ground-based detectors. Through such studies, it is also expected to shed lights on the nature of DM halos located in galaxies. 
 
Finally, we would like to note that generalization of the above study to cases with rotating BHs is a tantalizing direction, which will be more relevant to realistic situations. Along this direction, there have been some pioneer works studying DM halos surrounding rotating BHs \cite{Ferrer:2017,Xu:2018,Xu:2020,Xu:2021}, many of them applied the Newman-Janis algorithm to generalize the static Schwarzschild-like BHs to stationary Kerr-like ones \cite{AA14,Liu20}, but most of them considered only simple density profiles with divergent cusps at the origin \cite{Zhao96}. How to include more general density profiles, especially those with necessary cutoff at inner boundary, as in our models, remains an open problem. Such concerning, on the other hand, underlines the value of the analytic results obtained here.

\section*{Acknowledgements}

Z. S. thanks the hospitality of the Department of Physics and Astronomy of Baylor University during his visit. A.W. is partially supported by the U.S. Natural Science Foundation (NSF) under Grant No. PHY2308845.

\end{document}